\documentclass[openacc]{rsproca_new}
\usepackage{textcomp}
\usepackage{tabulary}
\usepackage[utf8]{inputenc}

\begin{document}

\title{LOUPE: Observing Earth from the Moon 
to prepare for detecting life on Earth-like exoplanets}

\author{
D. Klindžić$^{1,2}$, D.M. Stam$^{1}$, F. Snik$^{2}$, \\H.J. Hoeijmakers$^{3}$, M. Willebrands$^{2}$, \\T. Karalidi$^{4}$, V. Pallichadath,$^{1}$\\C.N. van Dijk$^{5}$, M. Esposito$^{5}$}

\address{$^{1}$ Aerospace Engineering, 
                Delft University of Technology,
                Kluyverweg 1, 2629 HS, Delft,
                The Netherlands \\
$^{2}$Leiden Observatory, Leiden University, P.O. Box 9513, 2300 RA Leiden, The Netherlands\\
$^{3}$ University of Bern, Center for Space and Habitability (CSH), Gesellschaftsstrasse 6 (G6), CH-3012, Bern, Switzerland\\
$^{4}$UCF Department of Physics, 4111 Libra Drive, Physical Sciences Bldg. 430, Orlando, FL\\
$^{5}$cosine Remote Sensing, Oosteinde 36, 2361 HE Warmond, The Netherlands\\}

\subject{astronomical polarimetry, spectropolarimetry, exoplanets}


\corres{Dora Klindžić\\
\email{d.klindzic@tudelft.nl}}

\begin{abstract}
LOUPE, the Lunar Observatory for Unresolved Polarimetry of the Earth, 
is a small, robust spectro-polarimeter for observing the Earth as an 
exoplanet. Detecting Earth-like planets in stellar habitable zones is 
one of the key challenges of modern exoplanetary science.
Characterising such planets and searching for 
traces of life requires the direct detection of their signals. 
LOUPE provides unique spectral flux and 
polarisation data of sunlight reflected by  
Earth, the only planet known to harbor life.
This data will be used to test numerical codes 
to predict signals of Earth-like exoplanets, 
to test algorithms that retrieve planet properties, 
and to fine-tune the design and
observational strategies of future space 
observatories. From the Moon, LOUPE will continuously see  
the entire Earth, enabling it to monitor the signal 
changes due to the planet's daily rotation, weather patterns, 
and seasons, across all phase angles. 
Here, we present both the science case and the 
technology behind LOUPE's instrumental and mission 
design.
\end{abstract}

\begin{fmtext}
\end{fmtext}

\maketitle

\section{Introduction}
\label{sect:intro}

Since the first discoveries of planets orbiting other stars in the 1990's, exoplanetary research has expanded explosively. 
Today, we know of more than 4,000 such worlds, ranging from gas giants more massive than Jupiter to rocky, terrestrial-type planets considered to be candidates for harboring life. 
Although we now know planets orbiting other stars are not uncommon, the occurrence rate of Earth-like planets in the habitable zone of Sun-like stars remains a highly debated topic. 
Statistical analysis of existing exoplanetary catalogs has shown that somewhere between 2-60\% of Sun-like stars may harbor planets similar to the Earth (or super-Earth) in their habitable zones\cite{nasaarchive}.
The exoplanetary catalogs are being expanded daily through space missions such as NASA's TESS (Transiting Exoplanetary Survey Satellite), which is expected to detect more than 14,000 exoplanets, of which over 2,100 will be smaller than 4 Earth-radii\cite{tess}; whereas ESA's upcoming PLATO (PLAnetary Transits and Oscillations of stars) mission will focus on habitable worlds around Solar-type stars, aiming to yield between 6 and 280 Earth analogs out of $\sim$4,600 total detections\cite{plato}.
Knowing that Earth-like rocky exoplanets might be more common than previously thought, the next step is investigating their atmospheres, surfaces, and biomarkers.
Although transit spectroscopy is a well-established method for 
characterizing gas giants, transit signals of the thin atmospheres of Earth-like exoplanets around Sun-like stars are undetectable
with present technology. A significant upcoming technological milestone 
for astronomy will be to achieve direct imaging,
in which a planet's  (reflected) starlight is observed separately from the 
light of its host star. 
Resolving the planet from its star will offer an opportunity to investigate 
its properties via spectral flux and polarization measurements.
Furthermore, direct imaging will enable non-transiting planets to be detected
and characterized.

The most reliable benchmark for characterizing Earth-like exoplanets is, naturally, the Earth. 
By placing the observer at a distance such that the Earth appears as an unresolved ``pale blue dot'', we may simulate  the observation of Earth as an exoplanet. 
In this single "dot", all the spectropolarimetric information from sunlight reflected off of Earth's oceans, continents, biomarkers and clouds is integrated into a spatially unresolved point.
If we can reliably extract this information from the unresolved signal and reverse-engineer the properties of the Earth as we know it, we will have developed a powerful tool for characterizing exoplanets, including their oceans, continents, atmospheric composition and life signatures, even if we are unable to spatially resolve them.

The Earth is already continuously being observed by remote-sensing satellites, which monitor e.g.\ atmospheric trace gas 
concentrations, crop health, and weather patterns.
Apart from the fact that there are currently no Earth 
remote-sensing satellites with polarimetric 
capabilities\footnote{NASA's PACE mission, which is planned
for launch at the end of 2022, will carry two polarimeters:
SPEXone and HARP2.}, such Low Earth Orbit (LEO) observations typically have their field-of-view limited to localized portions of the Earth's surface, and not the entire Earth's disc. 
A mosaic of such observations does not realistically 
represent the instantaneous single-pixel view of Earth, 
because the individual segments vary in terms of local time and 
weather conditions, 
and the distribution of local illumination and
viewing geometries is very different from the distribution 
when the Earth is viewed from afar. In particular, most 
satellites have a nadir viewing direction, and are in 
sun-synchronous orbits, observing a given location on 
Earth at more or less the same time of day.
Especially the polarization is very sensitive to the 
illumination and viewing angles\cite{hansentravis}.
Even satellites locked in geostationary orbit would not provide us with complete insight, as they only observe a single hemisphere, 
thus missing out on the variations due to the daily rotation.

Various reasons make observing the Earth from the Moon, 
from a Lunar orbit, or the Earth-Moon L1-point, 
rather than a low Earth orbit, crucial to the experiment: \\
\hspace*{0.5cm} 1. The Moon is sufficiently far away to allow a
spatially unresolved view of the whole Earth.\\
\hspace*{0.5cm} 2. For a lander on the Lunar surface, 
the Earth is always visible in a confined area in the sky.\footnote{Accounting for the approximate 8$^\circ$ apparent motion of the Earth on the celestial sphere due to the Lunar libration.} \\
\hspace*{0.5cm} 3. From the Moon, the Earth can be observed at all 
phase angles\footnote{The range of phase angles 
$\alpha$ an exoplanet can attain is
$90^\circ \leq \alpha \leq 90^\circ + i$, with orbital inclination angle $i$ equal to $0^\circ$ (90$^\circ$) for a face-on (edge-on) orbit.} 
during a month. \\
\hspace*{0.5cm} 4. From the Moon, the Earth's daily rotation can
be captured.\\
The latter provides a view of the entire Earth, 
and allows detecting changes in Earth's spectropolarimetric signals 
as continents and oceans rotate in and out of view.
Observations covering several months could reveal seasonal changes.
In light of these advantages, we propose the Lunar Observatory 
for Unresolved Polarimetry of Earth (LOUPE) \cite{karalidi}, 
a compact and small spectropolarimeter based on pioneering 
liquid crystal polarization optics, to accompany an orbiting, landing or roving mission on the near side of the Moon.  
LOUPE's tentative instrument design is presented in Section 
\ref{sect:design}, and the performance of a previous design iteration was validated in \cite{hoeijmakers}.

The main driver of LOUPE is to perform a long-term observing campaign of the Earth as if it were a spatially unresolved exoplanet, both in flux and polarisation, in order to provide the ongoing search for Earth-like exoplanets with the benchmark of an archetypal Earth.
Another approach to obtain such flux and
polarization data 
is through so-called ``Earthshine'' observations \cite{earthshine1, earthshine2, earthshine3, earthshine4},
where ground-based telescopes are used to search for back-scattered light of the Earth 
on the shadowed crescent of the lunar disk.
Although some spectral features of the Earth's flux were reported, such as the O$_2$-A band, the Vegetation Green Bump and Red Edge (VRE), this method is severely hampered by unknown depolarization effects of the reflection of polarized light by the Lunar surface, degradation of the signal as it re-enters the Earth's atmosphere to reach the observer, and the severe difficulties in monitoring the daily rotation and a broad range of phase angles. 
LOUPE would eliminate these problems by observing from the Moon itself, creating a dedicated spectropolarimetric observing platform with superior science return. 

LOUPE's aim is to pioneer spectropolarimetry as a uniquely qualified tool for exoplanet characterisation.
For ground-based telescopes, polarimetry makes it possible to differentiate between the reflected flux of a planet and the overpowering flux of its parent star\cite{exopolarimetry} even when direct detection is not possible from intensity alone, enabling us to find exoplanets which would otherwise be lost in the stellar glare.
This ability stems from the fact that sunlight and--more generally--light of Solar-type stars can be assumed to be unpolarized when averaged over the stellar disc, whereas the light scattered in a planetary atmosphere and/or reflected by a planetary surface will generally become polarized (up to 10\%.) 
Thus, measuring the polarized flux of a planet can be used to enhance the contrast between the two. 
Future space telescopes like HabEx/LUVOIR aim to deliver the intrinsic 10$^{-10}$ contrast to directly image an Earth orbiting a Solar-type star, and then polarimetry can be applied to further characterize the planet.
As Solar System observations have indicated\cite{venus}, the phase angle dependence of the linearly polarized spectrum of a planet is highly sensitive to atmospheric constituents and clouds, as well as surface features like vegetation, water, ice, snow or deserts (see Sect. \ref{sect:signals}.)
Therefore, the main advantage of spectropolarimetry is the ability to deliver unambiguous characterization of exoplanets, breaking the retrieval degeneracies arising from flux measurements alone.
In this way, polarimetry promises to reveal not only a plethora of new worlds, but also a plethora of new geomorphologies and biospheres.

One of our main goals of monitoring the Earth from afar is to gather benchmark data to test the radiative transfer codes which are being used to compute signals of rocky exoplanets\cite{stam2008,rossi}.
Such signals are crucial for the design of future (space) telescopes dedicated to the characterization of Earth-like exoplanets and for the development of algorithms to retrieve exoplanet characteristics\cite{aizawa}.
Previous attempts to study Earth as an exoplanet involved serendipitous measurements from deep space instruments used outside their intended mode of operation, e.g.\ the Galileo\cite{galileo}, Deep Impact\cite{impact}, Venus Express\cite{vex},  DSCOVR\cite{dscovr,nistar,epic}, and LCROSS\cite{lcross} space probes.
With their limited coverage, these experiments are unsuited to study the whole phase angle
range and to achieve full global coverage, nor could they measure polarisation, thus falling short of a complete and thorough characterization of the Earth's disk-integrated signal.
LOUPE's monitoring of the total flux that the Earth reflects
would also be valuable for climate research, as this reflected flux
and its spectral and temporal variations give insight into the 
amount of Solar energy that the Earth absorbs over time. 
The polarization signal of the Earth as a whole could provide new
information on high-altitude aerosol particles that contribute to the
Earth's radiation balance by reflecting incoming sunlight and by heating
up their ambient environment, as well as playing important roles
in chemical reactions \cite{sparc2006,kremser2016}.

Section~\ref{sect:signals} discusses interesting features in the 
Earth's flux and polarisation signals. Science requirements and the 
resulting instrument requirements and goals are presented in 
Sect.~\ref{sect:requirements}, and an overview of LOUPE's instrument 
design in Sect.~\ref{sect:design}. Conclusions are presented
in Sect.~\ref{sect:conclusions}.

\section{Earth's flux and polarisation signals}
\label{sect:signals}

In the absence of real spectropolarimetric data of the spatially unresolved 
Earth, we use numerical  simulations for the design of LOUPE.  
A key part of LOUPE's mission is to provide benchmark data for the 
improvement and refinement of such numerical simulations of exoplanet signals. 

We describe light as a Stokes (column) vector 
\cite{hansentravis}:
\begin{equation}
    \mathbf{F} = \left[
        F, Q, U, V \right],
        \label{eq1}
\end{equation}
with $F$ the total flux, $Q$ and $U$ the linearly polarised fluxes, 
and $V$ the circularly polarised flux (all in W~m$^{-2}$~nm$^{-1}$).
Fluxes $Q$ and $U$ are defined with respect to a reference
plane, for which we use the planetary scattering plane, 
i.e.\ the plane through the centers of the Sun, the Earth,
and the observer, which in our case is LOUPE on a Lunar orbiter or lander.
The degree of (linear) polarisation of the light is defined as
\begin{equation}
    P_L = \sqrt{Q^2 + U^2}/F.
    \label{eq2}
\end{equation}
The angle of polarisation, $\chi$, is also defined with 
respect to the reference plane:
    $\tan 2\chi = U/Q,$
where the sign of $\chi$ is such that $0 \leq \chi < \pi$ and that it equals that of $Q$\cite{hansentravis}.

\begin{figure}[!t]
\centering\includegraphics[width=\textwidth]{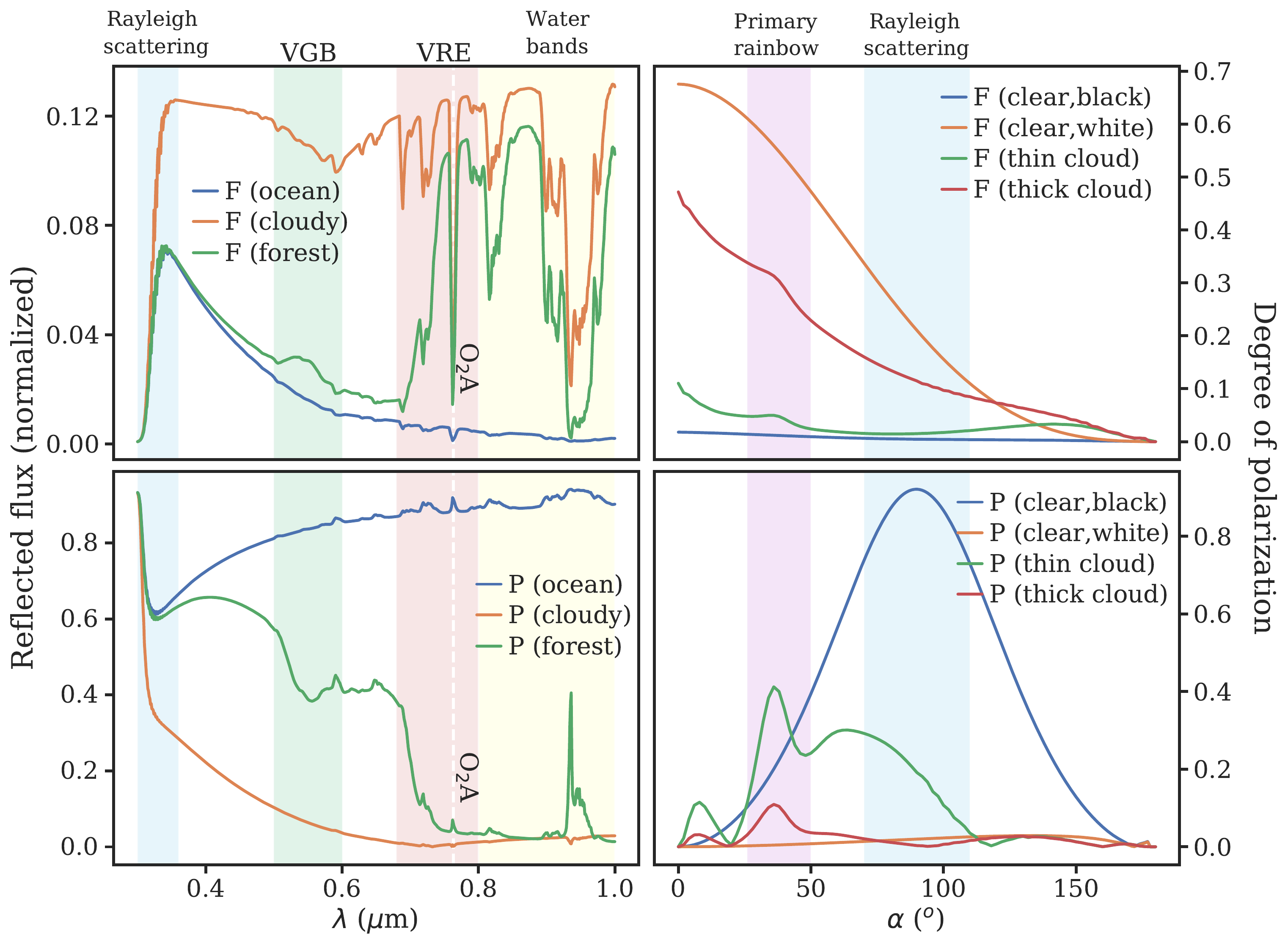}
\caption{Simulated reflected 
         flux and polarization at a phase angle $\alpha$ of 90$^\circ$
         as functions of the
         wavelength (left, from \cite{stam2008}) and as functions
         of $\alpha$ (right, from\cite{fauchez}) for horizontally
         homogeneous planets with varying surface properties and 
         cloud covers. The fluxes have been normalized 
         such that at $\alpha=0$, they equal the planet's 
         geometric albedo. Apart from absorption
         by oxygen (O$_2$), absorption by ozone and water-vapour
         is included.}
\label{fig:phasecurve}
\end{figure}

We compute ${\bf F}$ (Eq.~\ref{eq1}) of the visible and illuminated disk 
of a model Earth at a given phase angle $\alpha$ and wavelength $\lambda$
by dividing the disk into pixels with specific surface-atmosphere models 
(e.g. ocean-cloudy, forest-clear, desert-clear, etc.), computing the 
reflected Stokes vector for each pixel using an adding-doubling radiative
transfer algorithm \cite{groot2020,rossi,stam2008}, and summing up the local
vectors to obtain the disk-integrated, planetary Stokes vector.
An example of the simulated flux and polarization for an unresolved planet 
is shown in Fig.~\ref{fig:phasecurve}.

\begin{figure}[!t]
\centering\includegraphics[width=\textwidth]{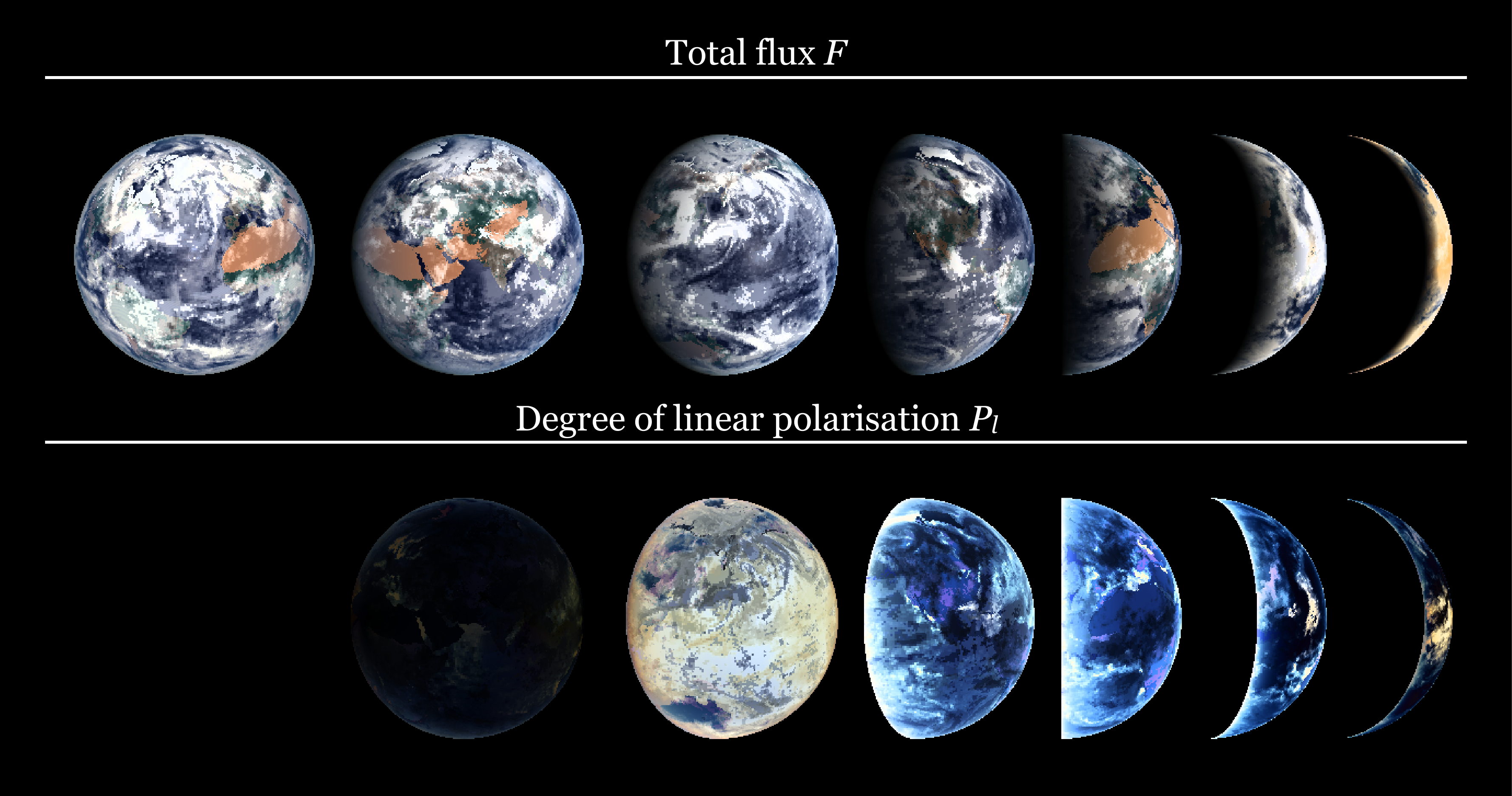}
\caption{Computed $F$ (top) and $P_L$ (bottom)
         for the Earth at phase angles $\alpha$ 
         starting with 0$^\circ$ on the left.
         An RGB-coloring scheme combined with a gray-scale was used
         to show both the spectral dependence and
         the absolute value of the reflected flux and the 
         polarization. Note that at 
         $\alpha=0^\circ$, $P_L$ is virtually zero.
         For these images, data described in \cite{groot2020} was used.}
\label{fig:earth}
\end{figure}

Figure~\ref{fig:earth} shows the computed $F$ and $P_L$ (Eq.~\ref{eq2}) 
for a spatially resolved model Earth at several phase angles.
Various features in $P_L$ stand out. Firstly, at $\alpha=0^\circ$, 
$P_L$ is zero across the disk because of the symmetric, back-scattering 
geometry for each pixel. Secondly, clouds generally have low $P_L$, 
and oceans with only Rayleigh-scattering gas above them, a high $P_L$.
Thirdly, at large phase angles, $P_L$ is highest in 
the red, due to the glint on the ocean\cite{wavyoceans}.
In the following subsections, we discuss some notable
reflective properties of the planetary surface and atmosphere.

\subsection{Continents and oceans}
\label{subsub:continents}

Due to plate tectonics, the Earth's surface is covered by continents and 
oceans. The continents are covered by various surface types, such as rocks,
sand, snow, and vegetation, each with characteristic (wavelength dependent) 
albedos and bidirectional reflectance functions (i.e.\ the angular 
distribution of the reflected total and polarized fluxes). 
Rough surfaces reflect more or less isotropically and strongly depolarize 
the incident light. Generally, the higher the albedo of such surfaces, 
the lower $P_L$, as the total flux increases but the polarized
fluxes do not \cite{stam2008}. 

Ocean surfaces exhibit specular, Fresnel, reflection, which is anisotropic 
and polarizing. The glint of sunlight on water is a particularly striking
feature, arising when the angles of incidence and reflection are equal.  
Waves influence the appearance of the glint: generally,
the higher the wind velocity, the higher the waves, 
and the broader the glint pattern that is expected to 
appear on the disk. With LOUPE, can investigate
this relationship and the influence of other wave parameters,
such as white caps and wave direction. 
Numerical results suggest that an ocean on a planet can uniquely be 
identified by a color change of $P_L$ of a planet at intermediate to large
phase angles: only with an ocean, a planet will change from blue, through
white, to red, with increasing $\alpha$, when observed using 
polarimetry\cite{wavyoceans}.

The numerically predicted effects of various reflecting surfaces on the
planetary phase curve both in total flux and polarization can be observed 
by LOUPE. 

\subsection{Vegetation}

Earth's vegetation owes its green color to a decrease in the absorption by 
chlorophyll (and thus an increase in the albedo) around $\lambda=500$~nm. 
However, as is evident from Fig.~\ref{fig:phasecurve}, the most distinct 
spectral feature of vegetation is not this 'Green Bump', but the dramatic 
brightness just outside the human visible range, the 'Vegetation Red Edge' 
(VRE)\cite{oxygen}.
Light-harvesting vegetation on exoplanets may have 
similar reflectance properties, as the VRE is hypothesized to limit excessive 
absorption of light at wavelengths where photosynthesis is inefficient. 
The VRE of exo-vegetation might cover different wavelengths than terrestrial
vegetation, but strong spectral features of unknown geological or atmospheric
origins could be worth investigating as possible biosignatures of alien 
vegetation.

A worthwhile exercise is to attempt to extract the VRE from the signal of a 
spatially unresolved Earth, as seen by LOUPE.
Simulations suggest that the VRE ought to be detectable even through optically 
thick clouds, and that its signature in polarization is even more pronounced, 
as it's located in a wavelength region where the degree of polarization is 
highly sensitive to the surface albedo\cite{stam2008}.
A tentative confirmation of the VRE in polarized light has been shown in 
Earthshine observations\cite{earthshine1}. 
LOUPE strives to confirm and improve the detection with its beneficial vantage 
point on the Moon, without the strongly depolarizing influence of a reflection 
by the Lunar surface.

Vegetation has also been shown to exhibit a small, but unambiguous circular 
polarization signature, as a consequence of the homochiral configuration of 
organic matter\cite{lsdpol}. 
A potential future ``super-LOUPE'' could be upgraded to perform full-Stokes 
demodulation, for example building on the design of the Life 
Signature Detection polarimeter (LSDpol\cite{lsdpol2}; 
see also \cite{sparks}), and retrieve the circularly polarized flux as an
additional biomarker to be studied.

\subsection{Clouds}

Clouds generally decrease $P_L$ because they add total flux but little
polarized flux. However, the phase angle variation of $P_L$ of a cloudy 
planet shows various interesting features, such as glories and, most notably,
rainbows. Rainbows are a well-known optical phenomenon formed when light is
scattered by airborne water droplets, such as rain droplets and also cloud 
droplets. In particular, the primary rainbow results from light rays which 
have undergone a single reflection inside spherical droplets. 
This rainbow exhibits dramatic peaks in both $F$ and $P_L$, as shown in 
Fig.~\ref{fig:phasecurve}.
Due to the small size of terrestrial cloud particles (as
compared to rain droplets), a cloudy Earth will
only show a significant rainbow peak in $P_L$\cite{karalidi:rainbows}.
The rainbow angle depends on the particle 
composition: with water clouds, these peaks would appear around 
$\alpha=40^\circ$, and could be used to identify the presence of liquid 
water clouds on exoplanets, even with small cloud coverage fractions and
partly overlaid with ice clouds\cite{karalidi:rainbows}.
Clouds with different compositions, such as sulfuric acid clouds which are 
present on Venus, would produce rainbows at different phase angles 
\cite{venus}.
Other numerical simulations show that the variability of $P_L$ of an 
exoplanet would reveal the spatial distribution of clouds\cite{rossi}.

Observing the Earth's clouds with LOUPE will give us a better understanding 
of the spectral and temporal variations in $F$ and $P_L$, which could be 
used to characterize the composition, spatial coverage and altitude of the
cloud cover on exoplanets.

\begin{figure}[!t]
\centering\includegraphics[width=\textwidth]{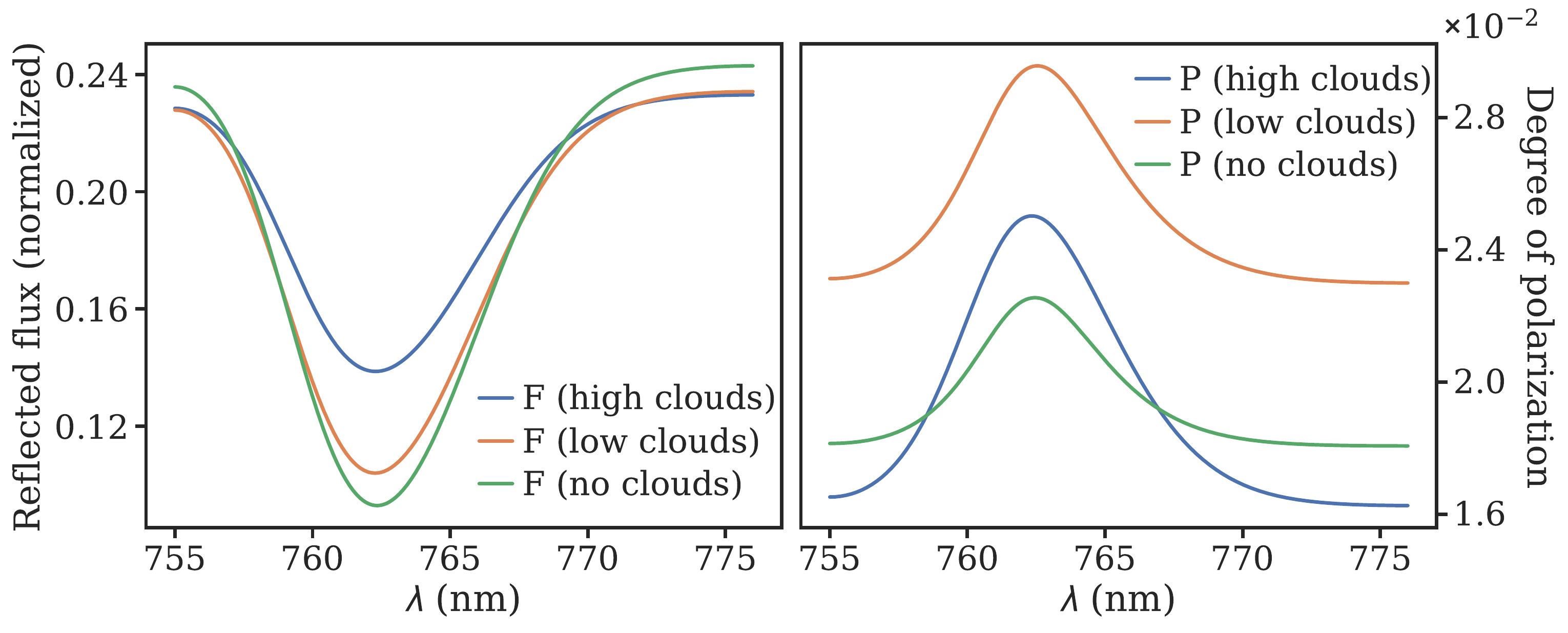}
\caption{Total flux $F$ (left) and degree of linear polarisation $P_L$ 
(right) as functions of $\lambda$ across the O$_2$ A-band for a model 
planet with a surface albedo of 0.6, without a cloud (green lines) or 
with a cloud with optical thickness 5.0 at different altitudes. 
The atmosphere consists of 5 layers with gaseous scattering optical 
thicknesses equal to (bottom to top): 0.01, 0.003, 0.005, 0.005, 0.002. 
The 'low cloud' (orange line) is in layer 3,
and the 'high cloud' (blue line) in layer 5.}
\label{fig:o2a}
\end{figure}

\subsection{Oxygen and trace gases}

The presence of abundant atmospheric oxygen (O$_2$) in thermodynamic 
disequilibrium is thought to be a robust biosignature, as on Earth, the
dominant source of O$_2$ is oxygenic photosynthesis \cite{oxygen}. 
Thus, an O$_2$-rich atmosphere could indicate the presence of 
photosynthetic organisms. An exoplanet's O$_2$ mixing ratio could be derived
from the depth of absorption bands in $F$ and $P_L$ spectra, of which the
A-band, centered around 760~nm is the least contaminated by absorption lines 
of water \cite{stam2008}. This depth, however, also depends on the presence 
of clouds: on Earth, measurements of $F$ across the A-band are routinely 
used to determine cloud-top altitudes \cite{o2acloudtop1,o2acloudtop2}, 
as the band depth increases with the amount of O$_2$ above the clouds, 
and thus with a decreasing cloud-top altitude.
This is evident in the $F$ and $P_L$ spectra across the band shown in 
Fig.~\ref{fig:o2a}, computed according to \cite{fauchez} and convolved 
with a Gaussian of 5~nm FWHM, informing LOUPE's goal instrument response
function (see Sect.~\ref{sect:design}.)  

The lines in Fig.~\ref{fig:o2a} were computed for a planet with a surface 
albedo of 0.6, completely covered by a cloud of optical thickness 5.0, and 
seen at $\alpha=60^\circ$. It is clear that the lower the cloud, the deeper 
the band (with respect to the continuum) in $F$, as the more absorbing gas 
is above it.
Because of the small atmospheric gaseous scattering optical thickness at 
these wavelengths, the continuum $F$ is insensitive to the cloud top altitude.
Note that without a cloud, the continuum $F$ is higher because the cloud 
particles are strongly forward scattering, thus scattering light towards 
the surface. $P_L$ is higher in the band because absorption suppresses 
(depolarizing) multiple scattering, and because it increases the average
scattering altitude and thus the scattering by the gas, which yields a 
higher $P_L$ at most scattering angles \cite{fauchez}.

LOUPE's observations will allow us to study the spectral and temporal
variations of Earth's $F$ and $P_L$ across the O$_2$ A-band and other
absorption bands, such as those of the trace gases ozone 
(O$_3$) and water vapour (H$_2$O), which are also indicative for 
a planet's habitability, across a range of phase angles and with that,
provide valuable insight into the diagnostic value of gaseous absorption 
bands in exoplanet spectra.

\section{Scientific and technical requirements}
\label{sect:requirements}

As outlined above, the top-level science requirements for LOUPE are:
\vspace*{-0.3cm}
\begin{itemize}
    \item Perform near-instantaneous (snapshot) spectropolarimetry of the entire Earth.
    \item Detect the presence of liquid water oceans and clouds.
    \item Derive and monitor atmospheric properties, e.g. via Rayleigh scattering.
    \item Detect the O$_2$-A band in $F$ and $P_L$, and its variance with cloud cover and altitude, and $\alpha$.
    \item Detect the Chlorophyll Green Bump and Vegetation Red Edge, the spectroscopic signature of plant life.
    \item Derive a map of continents from the disk-integrated signal and identify notable features, such as rain forests, deserts and ice caps.
\end{itemize}{}
\vspace*{-0.3cm}
LOUPE shall perform its science goals by recording and demodulating the 
disk-integrated Stokes vector of sunlight reflected from the Earth.
The minimum and goal technical requirements for LOUPE's mission are stated 
in Table \ref{tbl:requirements}.

In order to maximize deployment opportunities, LOUPE will be
prototyped with platform versatility in mind, so that it may be suitable 
for multiple use cases (including geostationary, Lunar orbiting and landing
scenarios).
To keep a first proof-of-concept version of LOUPE as simple as possible, 
key trade-offs may be undertaken, such as limiting the instrument 
not to measure the circular polarization $V$, but only $F$, $Q$, and $U$. 
This trade-off is justified by the fact that, although $V$ has the potential 
to be considered a biosignature of homochiral life \cite{lsdpol}, 
$V$ of light reflected by an (exo)planet is several orders of magnitude 
lower than the linearly polarized flux \cite{kemp,rossiV,groot2020} and
neglecting it introduces no significant errors in $F$, $Q$, and $U$ 
\cite{stamv}. 
Some other capabilities of LOUPE are functionally optional as 
well, such as radiometry, though its addition could provide data for climate 
research. 
The additional capability to resolve the Earth at the continent scale would 
also introduce additional signal processing difficulties, as the ability to 
perform straightforward disk-integration of the signal is integral to LOUPE's 
mission.
Weighing the advantage of a wide field of view with passive pointing against a 
narrower, baffled field of view with better protection from the Solar glare, 
but a need for active pointing, is another topic for contemplation in the 
design process.

\begin{table}[!b]
    \centering
    \scriptsize  
    \settowidth\tymin{\textbf{Requirement}}
    \setlength\extrarowheight{3pt} 
    \begin{tabulary}{\linewidth}{L|C|C|C}
        \textbf{Requirement} & \textbf{Minimum} & \textbf{Goal} & \textbf{Rationale} \\
        \hline
        Spectral range & 500-800~nm & 400-1000~nm & (Rayleigh scattering $\sim$400~nm,) Green bump: 500-600~nm, VRE: 700-760~nm, O$_2$A band: 750-770~nm, (H$_2$O bands: 800-1000~nm)\\
        Spectral resolution & 20~nm & 5~nm & 20~nm suffices for broad spectral features, 5~nm would enable a detailed look into the O$_2$A band (possible implementation as a separate channel).\\
        Polarization parameters & $F(\lambda),P_L(\lambda),$ $\chi_L(\lambda)$ & full Stokes & Circularly polarised flux $V$ could be introduced in a separate channel, using modulation optics similar to LSDpol\cite{lsdpol2}. \\
        Polarimetric sensitivity & 0.1\% & <10$^{-4}$ & The smallest detectable change in fractional polarization. Must be able to detect the weaker features in Fig.~\ref{fig:phasecurve}, e.g. Green Bump, O2A band\cite{stam2008}. \\
        Polarimetric accuracy & 1\% & <10$^{-3}$ & Accuracy of measurements limited by systematic errors. Planetary properties affect the polarization signal at below 1\% level\cite{groot2020}. \\
        Relative photometry & 3\% & 1\% & To track the diurnal changes in flux with a sufficient SNR\cite{karalidi}. \\
        Radiometric accuracy & No requirement & $\sim$1\% & Retrieval accuracy for the absolute value of flux. Goal derived from NISTAR radiometric instrument aboard the DSCOVR mission for potential climate research applications\cite{nistar}. \\
        Field of view / pointing & 20$^o\times$20$^o$, rough pointing & Avoidance of horizon \& Sun, active pointing & Diameter of Earth is 2$^o$, Lunar libration $\pm8^o$. In case of polar lander, ensure Earth is in FOV \textit{a priori} in the passive pointing scenario, \textit{a posteriori} with active pointing. \\
        Spatial sampling & Unresolved & Continent-sized & Measurements must enable easy disk-integration of signal. \\
        Temporal sampling & Hourly & Multiple/hour & To capture the Earth's diurnal rotation and enable continent mapping. Minimum derived from EPIC imager aboard the DSCOVR mission\cite{epic}.  \\
        Mission duration & 1 month & 1+ years & To get an overview of phase angles, and additionally the seasonal variation in polarised flux. \\
        Mass & <1~kg & 300~g & Ensure minimal addition to payload mass. Includes electronics and protective mechanism. \\
        Volume & 1~U & <1~U & Ensure versatility for potential 1U CubeSat proof of principle. \\
        Moving parts & Protective lid (single-use) & Active pointing \& protection & Active protection from the Solar glare or Lunar dust would improve data quality over data-driven masking. \\
        Data bandwidth & $\sim$50~MB/day & >100~MB/day & One compressed observation is expected to produce 2~MB of data. Temporal sampling will determine total daily load.
    \end{tabulary}
    \caption{Minimum and goal technical requirements for LOUPE.}
    \label{tbl:requirements}
\end{table}

\section{LOUPE instrument design}
\label{sect:design}

\begin{figure}[!t]
\centering\includegraphics[width=5in]{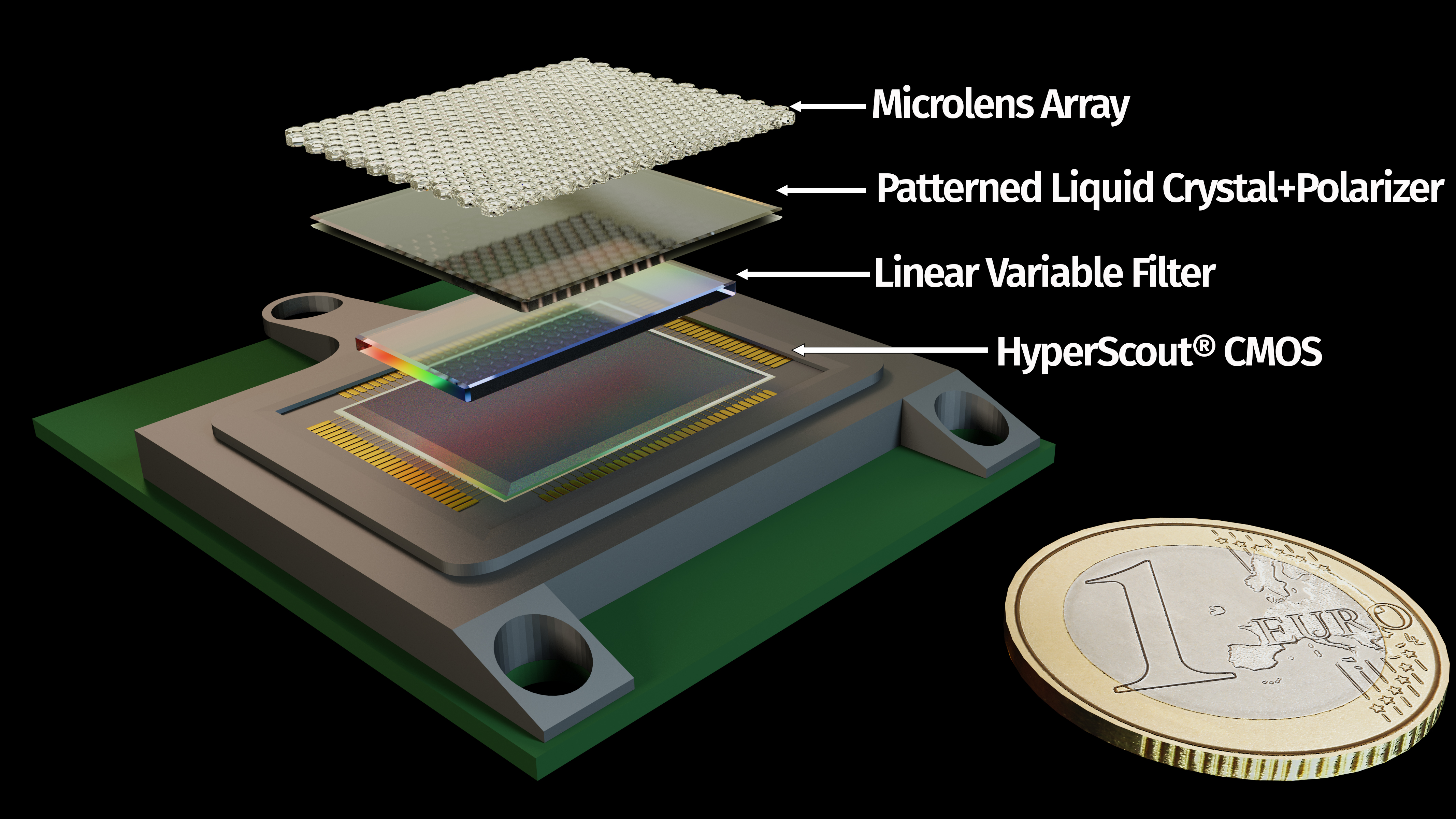}
\caption{A 3D render of the current LOUPE concept, 
         with a \texteuro1 coin for scale. 
}
\label{fig:3d}
\end{figure}

The leading instrument design principle adopted for LOUPE is to create a
compact, low-mass, low-volume, space-ready hyperspectropolarimeter with no
moving parts \cite{karalidi}. 
These constraints require creative solutions from the cutting edge of 
hyperspectral\footnote{Unlike traditional digital imaging, which records each 
pixel value in three discrete bands of red, green and blue visible light, 
spectral imaging records each pixel spectrum in a larger number of wavelength 
bands, possibly extending beyond the visible. Accordingly, hyperspectral 
imaging records spectra in continuous spectral bands with a very fine 
wavelength resolution.} and polarimetric instrument design, where polarimeters 
traditionally used active rotating optics (temporal modulation) or 
beam-splitting (spatial modulation) 
\cite{polarimeter1,polarimeter2,polarimeter3}.
Since the first design iteration and proof-of-concept study of LOUPE 
\cite{hoeijmakers}, further improvements forgo the use of bulky imaging and 
reimaging optics, resulting in a compact, solid-state instrument with a novel 
approach to snapshot spectropolarimetry. 
Figure~\ref{fig:3d} shows a tentative 3D-render of LOUPE's latest design.

\begin{figure}[!t]
\centering\includegraphics[width=\textwidth]{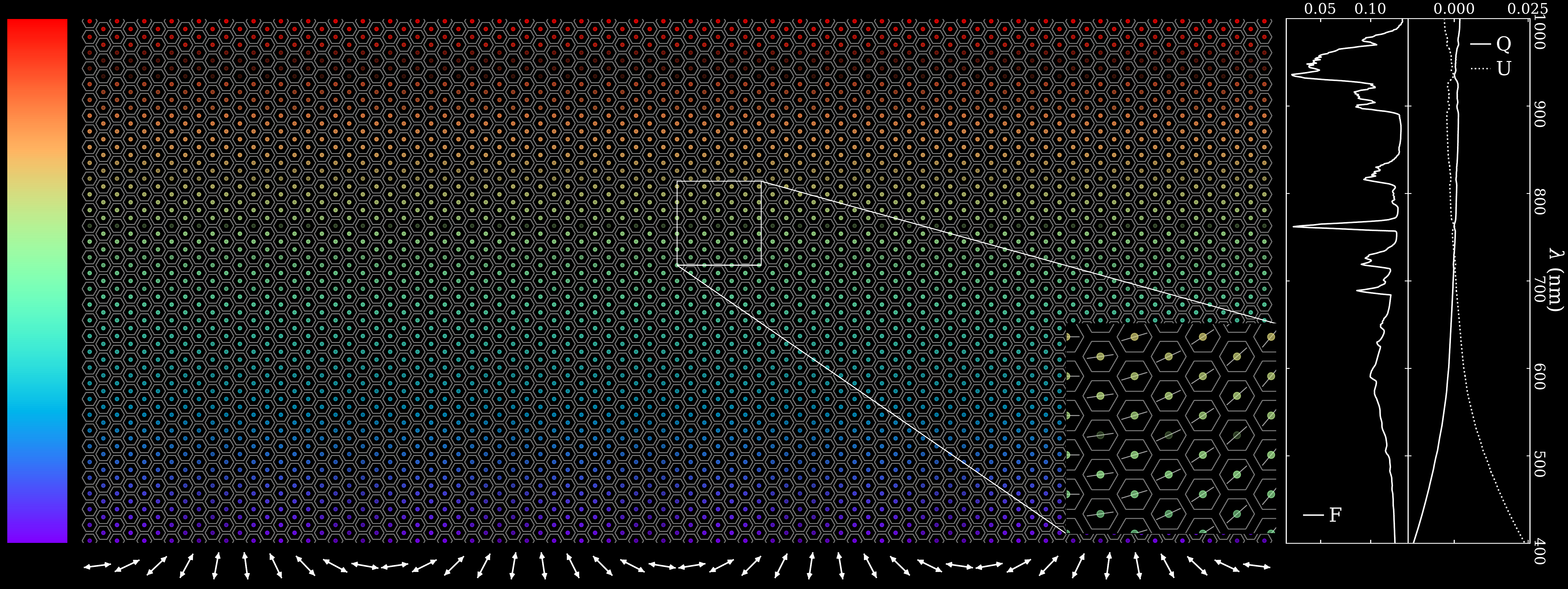}
\caption{A simulated LOUPE detector snapshot. Each colored dot is an 
         unresolved Earth-image, filtered spectrally along the vertical
         (denoted by color), with polarization modulation along the 
         horizontal (denoted by the arrows). The wavelength dependence of 
         $F$, $Q$ and $U$ is plotted on the right-hand side. The input 
         spectrum is a simulated fully cloudy planet, with the spectral 
         resolution set to $\sim$3~nm, and the detector is rotated by 
         30$^o$ with respect to the planetary scattering plane.}
\label{fig:dots}
\end{figure}

The challenge for LOUPE is collapsing four-dimensional data ($F,P,\chi$, and
$\lambda$) onto a two-dimensional detector, instantaneously for Earth's 
entire disk. 
The instrument will be built on cosine Remote Sensing's\footnote{\url{https://www.cosine.nl/}}  HyperScout\textregistered\footnote{\url{https://hyperscout.nl/}}\, hyperspectral imaging platform\cite{hyperscout1,hyperscout2}, space qualified and operating in Earth orbit for almost two and a half years\footnote{\url{http://www.esa.int/ESA_Multimedia/Images/2020/05/HyperScout_view_of_Netherlands}}, which is based on CMOS and linear variable filter (LVF) technologies.
Because apart from the Sun, the Earth is the brightest object in the sky as 
seen from the Moon, we can use a wide-field micro-lens array (MLA) instead 
of a traditional telescope objective system.
Each "fisheye" MLA-lenslet focuses the Earth as a dot on the detector (see 
Fig.~\ref{fig:dots}). 
We therefore forfeit the ability to resolve features on the Earth's disk, in 
favour of recording an unresolved point source, similar to observations of 
distant exoplanets.

By overlaying a linear variable filter (LVF) atop the detector, every pixel 
will be filtered spectrally in the direction of the LVF gradient (see 
Fig.~\ref{fig:dots}). The polarization information is encoded in the
perpendicular direction (see Fig.~\ref{fig:dots}) using a technique 
of cross-spectral modulation analogous to the rotating retarder 
polarimeter \cite{sparks,lsdpol2}. Uniquely to this design, and similarly 
to that of LOUPE's 'cousin', LSDpol, this is achieved by placing a linear
polarizer and a patterned liquid crystal (PLC)\cite{plc} on top of the
spectrometer. The liquid crystal pattern is such that it behaves as an
achromatic half-wave plate\cite{plc2} for all wavelengths of interest. 
This combination of polarizer and PLC acts as a passive modulator 
superimposing a sinusoidal modulation on the flux spectrum in the 
cross-spectral direction. This modulation has the same form as the case 
of a rotating wave plate polarimeter described in \cite{polarimeter1}, 
and ensuring several modulation cycles across the detector provides 
redundancy in case of bad pixels or local dust accumulation.
The amplitude of this modulation scales with $P_L$, and its phase with $\chi$. 
By demodulating the signal in post-processing, the full polarization 
information can be retrieved in parallel with a spectral measurement at full 
spectral resolution. Additional resolution across spectral regions of 
interest, e.g.\ the O$_2$ A-band, can be achieved by installing ring 
resonators as bandpass filters on separate "pixels" next to the
HyperScout\textregistered\, focal plane.

In conclusion, a LOUPE observation will consist of an array of ``pale (blue) 
dots'' in all colors of the spectrum, modulated with respect to angle and 
degree of polarization. We can then extract the disk-integrated Stokes vector
by demodulating this two-dimensional array of dots, and proceed to compare 
it to our numerically simulated planet signals. The features we identify in 
our analysis can be verified by comparison to satellite data. 

Another benefit of this compact design is that the need for instrument 
pointing has been effectively removed.
The offset of an Earth-dot from its respective lenslet center in 
Fig.~\ref{fig:dots} is directly related to the incidence angle of the 
Earth-light, enabling the retrieval of Earth's position relative to the 
detector \textit{a posteriori}. 
This is crucial for both the spectral and the polarization pre-flight 
calibration, which strongly depend on the incidence angle.
In addition to the data-driven calibrations enabled
by LOUPE's elegant design, vicarious calibrations can be performed using, 
e.g.\ bright starlight of known properties.
Furthermore, any persistent features caught in the instrument's field of view 
-- such as the Lunar surface -- can be corrected for. 
As long as LOUPE has a direct line of sight to Earth, 
even accounting for Lunar libration, active mechanical pointing is not
required. Yet another benefit of the MLA-design is redundancy: bad pixels 
or lenses covered by Lunar dust can be corrected for in post-processing.

The preliminary design fits well within the dimensions of 1~U and ca.~300~g 
(Table~1), and can be adapted to a variety of landing, roving or orbiting 
missions. For instance, for a roving mission to the Lunar south pole, where 
the Earth remains close to the horizon, fold mirrors can be installed to 
ensure an image of the Earth is reflected onto the horizontal LOUPE detector
without actively pointing the instrument. Alternatively, installing multiple 
LOUPEs so that they face various directions and span the sky might be the 
solution for a lander or an orbiting platform such as the Lunar 
Gateway.\footnote{\url{http://www.esa.int/Science_Exploration/Human_and_Robotic_Exploration/Exploration/Gateway}}. 

As such, LOUPE's lightweight and robust design is a low-cost addition to any 
existing Lunar landing or roving mission with minimal impact to payload mass, 
power consumption and down-link load, as each hourly observation is expected 
to produce 2 MB of data at an estimated 1~kJ per image.

\section{Conclusion}
\label{sect:conclusions}

In the quest to characterize terrestrial exoplanets, the first step is an 
introspective look at Earth as our benchmark.
The Lunar Observatory for Unresolved Polarimetry of the Earth (LOUPE) applies 
pioneering hyperspectropolarimetric techniques to observe 
Earth as an exoplanet from the Moon. 
LOUPE's science mission is to guide future exoplanet observing campaigns by 
offering improved models of exoplanetary flux and polarization spectra, 
including the ability to recognize features such as clouds, continents, 
oceans, vegetation and oxygen abundance on worlds we cannot resolve past a 
single pixel. 
LOUPE's novel design is being prototyped around state of the art patterned 
liquid crystal optics for polarimetry, working in tandem with the cosine 
HyperScout\textregistered\, hyperspectral imager for spectroscopy, which was 
first launched to orbit in 2018. Following design, manufacturing, testing and 
calibration, the first flight qualified model of LOUPE is expected to be 
ready in 2022, resulting in a compact, light-weight addition to 
any mission orbiting or landing on the near side of the Moon. \vskip6pt

\enlargethispage{20pt}





\funding{We acknowledge funding through NSO/PIPP NSOKNW.2017.003. Frans Snik acknowledges the support of ERC StG 678194 FALCONER.}




\begin{thebibliography}{9}

\bibitem{nasaarchive} NASA Exoplanet Archive: List of Planet Occurrence Rate Papers, updated 18 May 2020. \url{https://exoplanetarchive.ipac.caltech.edu/docs/occurrence_rate_papers.html}

\bibitem{tess} 
Barclay, T., Pepper, J., \& Quintana, E. V. (2018). A Revised Exoplanet Yield from the Transiting Exoplanet Survey Satellite (TESS). The Astrophysical Journal Supplement Series, 239(1), 2. https://doi.org/10.3847/1538-4365/aae3e9

\bibitem{plato} ``PLATO – Revealing habitable worlds around solar-like stars.''\
Definition Study Report, ESA-SCI(2017)1, April 2017

\bibitem{hansentravis}
Hansen, J. E., \& Travis, L. D. (1974). Light scattering in planetary atmospheres. Space Science Reviews, 16(4), 527–610. https://doi.org/10.1007/bf00168069

\bibitem{karalidi} 
Karalidi, T., Stam, D. M., Snik, F., Bagnulo, S., Sparks, W. B., \& Keller, C. U. (2012). Observing the Earth as an exoplanet with LOUPE, the lunar observatory for unresolved polarimetry of Earth. Planetary and Space Science, 74(1), 202–207. https://doi.org/10.1016/j.pss.2012.05.017

\bibitem{hoeijmakers} 
Hoeijmakers, H. J., Arts, M. L. J., Snik, F., Keller, C. U., Stam, D. M., \& Kuiper, J. M. (2016). Design trade-off and proof of concept for LOUPE, the Lunar Observatory for Unresolved Polarimetry of Earth. Optics Express, 24(19), 21435. https://doi.org/10.1364/oe.24.021435

\bibitem{earthshine1} 
Sterzik, M. F., Bagnulo, S., Stam, D. M., Emde, C., \& Manev, M. (2019). Spectral and temporal variability of Earth observed in polarization. Astronomy \& Astrophysics, 622, A41. https://doi.org/10.1051/0004-6361/201834213

\bibitem{earthshine2} 
Takahashi, J., Itoh, Y., Akitaya, H., Okazaki, A., Kawabata, K., Oasa, Y., \& Isogai, M. (2013). Phase Variation of Earthshine Polarization Spectra. Publications of the Astronomical Society of Japan, 65(2), 38. https://doi.org/10.1093/pasj/65.2.38

\bibitem{earthshine3}
Sterzik, M. F., Bagnulo, S., \& Palle, E. (2012). Biosignatures as revealed by spectropolarimetry of Earthshine. Nature, 483(7387), 64–66. https://doi.org/10.1038/nature10778

\bibitem{earthshine4} 
Bazzon, A., Schmid, H. M., \& Gisler, D. (2013). Measurement of the earthshine polarization in the B, V, R, and I bands as function of phase. Astronomy \& Astrophysics, 556, A117. https://doi.org/10.1051/0004-6361/201321855

\bibitem{exopolarimetry} 
Keller, C. U., Schmid, H. M., Venema, L. B., Hanenburg, H., Jager, R., Kasper, M., Martinez, P., Rigal, F., Rodenhuis, M., Roelfsema, R., et al. (2010). EPOL: the exoplanet polarimeter for EPICS at the E-ELT. In I. S. McLean, S. K. Ramsay, \& H. Takami (Eds.), Ground-based and Airborne Instrumentation for Astronomy III. SPIE. https://doi.org/10.1117/12.857626

\bibitem{venus}  
Hansen, J. E., \& Hovenier, J. W. (1974). Interpretation of the Polarization of Venus. Journal of the Atmospheric Sciences, 31(4), 1137–1160. https://doi.org/10.1175/1520-0469(1974)031<1137:iotpov>2.0.co;2

\bibitem{stam2008} 
Stam, D. M. (2008). Spectropolarimetric signatures of Earth-like extrasolar planets. Astronomy \& Astrophysics, 482(3), 989–1007. https://doi.org/10.1051/0004-6361:20078358

\bibitem{rossi} 
Rossi, L., \& Stam, D. M. (2017). Using polarimetry to retrieve the cloud coverage of Earth-like exoplanets. Astronomy \& Astrophysics, 607, A57. https://doi.org/10.1051/0004-6361/201730586

\bibitem{aizawa} 
Aizawa, M., Kawahara, H., \& Fan, S. (2020). Global Mapping of an Exo-Earth Using Sparse Modeling. The Astrophysical Journal, 896(1), 22. https://doi.org/10.3847/1538-4357/ab8d30

\bibitem{galileo} 
Sagan, C., Thompson, W. R., Carlson, R., Gurnett, D., \& Hord, C. (1993). A search for life on Earth from the Galileo spacecraft. Nature, 365(6448), 715–721. https://doi.org/10.1038/365715a0

\bibitem{impact} 
Cowan, N. B., Agol, E., Meadows, V. S., Robinson, T., Livengood, T. A., Deming, D., Lisse, C. M., A’Hearn, M. F., Wellnitz, D. D., Seager, S., \& Charbonneau, D. (2009). Alien maps of an ocean-bearing world. The Astrophysical Journal, 700(2), 915–923. https://doi.org/10.1088/0004-637x/700/2/915

\bibitem{vex} 
Oliva, F., Piccioni, G., D'Aversa, E., Bellucci, G., Sindoni, G., Filacchione, G., \& Capaccioni, F.. (2017). Earth as an exoplanet: VIRTIS-M/Venus Express data analysis. European Planetary Science Congress, EPSC2017-531.

\bibitem{dscovr} 
Jiang, J. H., Zhai, A. J., Herman, J., Zhai, C., Hu, R., Su, H., Natraj, V., Li, J., Xu, F., \& Yung, Y. L. (2018). Using Deep Space Climate Observatory Measurements to Study the Earth as an Exoplanet. The Astronomical Journal, 156(1), 26. https://doi.org/10.3847/1538-3881/aac6e2

\bibitem{nistar} 
Carlson, B., Lacis, A., Colose, C., Marshak, A., Su, W., \& Lorentz, S. (2019). Spectral Signature of the Biosphere: NISTAR Finds It in Our Solar System From the Lagrangian L‐1 Point. Geophysical Research Letters, 46(17–18), 10679–10686. https://doi.org/10.1029/2019gl083736

\bibitem{epic}
Marshak, A., Herman, J., Adam, S., Karin, B., Carn, S., Cede, A., Geogdzhayev, I., Huang, D., Huang, L.-K., Knyazikhin, Y., Kowalewski, M., Krotkov, N., Lyapustin, A., McPeters, R., Meyer, K. G., Torres, O., \& Yang, Y. (2018). Earth Observations from DSCOVR EPIC Instrument. Bulletin of the American Meteorological Society, 99(9), 1829–1850. https://doi.org/10.1175/bams-d-17-0223.1

\bibitem{lcross} 
Robinson, T. D., Ennico, K., Meadows, V. S., Sparks, W., Bussey, D. B. J., Schwieterman, E. W., \& Breiner, J. (2014). Detection of Ocean Glint and Ozone Absorption Using LCROSS Earth Observations. The Astrophysical Journal, 787(2), 171. https://doi.org/10.1088/0004-637x/787/2/171

\bibitem{sparc2006} 
Thomason, L., \& Peter, Th. (editors). (2006).
SPARC Assessment of Stratospheric Aerosol Properties (ASAP).
SPARC Report No. 4,WCRP-124,WMO/TD-No. 1295, 322 pp.

\bibitem{kremser2016} 
Kremser, S., Thomason, L. W., von Hobe, M., Hermann, M., Deshler, T., Timmreck, C., Toohey, M., Stenke, A., Schwarz, J. P., Weigel, R., et al. (2016). Stratospheric aerosol-Observations, processes, and impact on climate. Reviews of Geophysics, 54(2), 278–335. https://doi.org/10.1002/2015rg000511

\bibitem{groot2020} 
Groot, A., Rossi, L., Trees, V. J. H., Cheung, J. C. Y., \& Stam, D. M. 
(2020).`Colors of an Earth-like exoplanet. Accepted for publication in Astronomy and Astrophysics.

\bibitem{fauchez} 
Fauchez, T., Rossi, L., \& Stam, D. M. (2017). The O2A-Band in the Fluxes and Polarization of Starlight Reflected by Earth-Like Exoplanets. The Astrophysical Journal, 842(1), 41. https://doi.org/10.3847/1538-4357/aa6e53

\bibitem{wavyoceans} 
Trees, V. J. H., \& Stam, D. M. (2019). Blue, white, and red ocean planets. Astronomy \& Astrophysics, 626, A129. https://doi.org/10.1051/0004-6361/201935399

\bibitem{oxygen} 
Meadows, V. S., Reinhard, C. T., Arney, G. N., Parenteau, M. N., Schwieterman, E. W., Domagal-Goldman, S. D., Lincowski, A. P., Stapelfeldt, K. R., Rauer, H., DasSarma, S., et al. (2018). Exoplanet Biosignatures: Understanding Oxygen as a Biosignature in the Context of Its Environment. Astrobiology, 18(6), 630–662. https://doi.org/10.1089/ast.2017.1727

\bibitem{lsdpol} 
Patty, C. H. L., ten Kate, I. L., Buma, W. J., van Spanning, R. J. M., Steinbach, G., Ariese, F., \& Snik, F. (2019). Circular Spectropolarimetric Sensing of Vegetation in the Field: Possibilities for the Remote Detection of Extraterrestrial Life. Astrobiology, 19(10), 1221–1229. https://doi.org/10.1089/ast.2019.2050

\bibitem{lsdpol2} 
Snik, F., Keller, C., Doelman, D., Kühn, J., Patty, L., Hoeijmakers, J., Pallichadath, V., Stam, D., Pommerol, A., Poch, O., \& Demory, B.-O. (2019). A snapshot full-Stokes spectropolarimeter for detecting life on Earth. In F. Snik, J. M. Craven, \& J. A. Shaw (Eds.), Polarization Science and Remote Sensing IX. SPIE. https://doi.org/10.1117/12.2529548

\bibitem{sparks} 
Sparks, W. B., Germer, T. A., \& Sparks, R. M. (2019). Classical Polarimetry with a Twist: A Compact, Geometric Approach. Publications of the Astronomical Society of the Pacific, 131(1001), 075002. https://doi.org/10.1088/1538-3873/ab1933

\bibitem{karalidi:rainbows} 
Karalidi, T., Stam, D. M., \& Hovenier, J. W. (2012). Looking for the rainbow on exoplanets covered by liquid and icy water clouds. Astronomy \& Astrophysics, 548, A90. https://doi.org/10.1051/0004-6361/201220245

\bibitem{o2acloudtop1} 
Fischer, J., \& Grassl, H. (1991). Detection of Cloud-Top Height from Backscattered Radiances within the Oxygen A Band. Part 1: Theoretical Study. Journal of Applied Meteorology, 30(9), 1245–1259. https://doi.org/10.1175/1520-0450(1991)030<1245:docthf>2.0.co;2

\bibitem{o2acloudtop2}
Fischer, J., Cordes, W., Schmitz-Peiffer, A., Renger, W., \& Mörl, P. (1991). Detection of Cloud-Top Height from Backscattered Radiances within the Oxygen A Band. Part 2: Measurements. Journal of Applied Meteorology, 30(9), 1260–1267. https://doi.org/10.1175/1520-0450(1991)030<1260:docthf>2.0.co;2

\bibitem{kemp}  
Kemp, J. C., Swedlund, J. B., Murphy, R. E., \& Wolstencroft, R. D. (1971). Physical Sciences: Circularly Polarized Visible Light from Jupiter. Nature, 231(5299), 169–170. https://doi.org/10.1038/231169a0

\bibitem{rossiV}
Rossi, L., \& Stam, D. M. (2018). Circular polarization signals of cloudy (exo)planets. Astronomy \& Astrophysics, 616, A117. https://doi.org/10.1051/0004-6361/201832619

\bibitem{stamv}
Stam, D. M., \& Hovenier, J. W. (2005). Errors in calculated planetary phase functions and albedos due to neglecting polarization. Astronomy \& Astrophysics, 444(1), 275–286. https://doi.org/10.1051/0004-6361:20053698

\bibitem{polarimeter1} 
Snik, F., \& Keller, C. U. (2013). Astronomical Polarimetry: Polarized Views of Stars and Planets. In Planets, Stars and Stellar Systems (pp. 175–221). Springer Netherlands. https://doi.org/10.1007/978-94-007-5618-2\_4

\bibitem{polarimeter2} 
Rodenhuis, M., Snik, F., van Harten, G., Hoeijmakers, J., \& Keller, C. U. (2014). Five-dimensional optical instrumentation: combining polarimetry with time-resolved integral-field spectroscopy. In D. B. Chenault \& D. H. Goldstein (Eds.), Polarization: Measurement, Analysis, and Remote Sensing XI. SPIE. https://doi.org/10.1117/12.2053241

\bibitem{polarimeter3} 
Snik, F., Craven-Jones, J., Escuti, M., Fineschi, S., Harrington, D., De Martino, A., Mawet, D., Riedi, J., \& Tyo, J. S. (2014). An overview of polarimetric sensing techniques and technology with applications to different research fields. In D. B. Chenault \& D. H. Goldstein (Eds.), Polarization: Measurement, Analysis, and Remote Sensing XI. SPIE. https://doi.org/10.1117/12.2053245

\bibitem{hyperscout1} 
Esposito, M., Conticello, S. S., Pastena, M., \& Carnicero Dominguez, B. (2019). In-orbit demonstration of artificial intelligence applied to hyperspectral and thermal sensing from space. In C. D. Norton, T. S. Pagano, \& S. R. Babu (Eds.), CubeSats and SmallSats for Remote Sensing III. SPIE. https://doi.org/10.1117/12.2532262

\bibitem{hyperscout2} 
Esposito, M., \& Zuccaro Marchi, A. (2019). In-orbit demonstration of the first hyperspectral imager for nanosatellites. In N. Karafolas, Z. Sodnik, \& B. Cugny (Eds.), International Conference on Space Optics — ICSO 2018. SPIE. https://doi.org/10.1117/12.2535991

\bibitem{plc} 
Miskiewicz, M. N., \& Escuti, M. J. (2014). Direct-writing of complex liquid crystal patterns. Optics Express, 22(10), 12691. https://doi.org/10.1364/oe.22.012691

\bibitem{plc2} 
Komanduri, R. K., Lawler, K. F., \& Escuti, M. J. (2013). Multi-twist retarders: broadband retardation control using self-aligning reactive liquid crystal layers. Optics Express, 21(1), 404. https://doi.org/10.1364/oe.21.000404
     
\end{thebibliography}
\end{document}